\documentclass[useAMS,usenatbib]{mn2e}
\usepackage {graphicx}
\usepackage {amssymb}
\title[Variabilities from Hyper-accretion Disks]
{Variabilities of Gamma-ray Bursts from Black Hole Hyper-accretion Disks}
\author[Lin et al.]
{\parbox{\textwidth}{Da-Bin Lin$^{1,2}$\thanks{E-mail: lindabin@gxu.edu.cn}, Zu-Jia Lu$^{1,2}$, Hui-Jun Mu$^{3}$, Tong Liu$^{3,5}$, Shu-Jin Hou$^{4}$, Jing L{\"u}$^{1,2}$,\\
Wei-Min Gu$^{3}$, and En-Wei Liang$^{1,2}$\\}\\
$^1$GXU-NAOC Center for Astrophysics and Space Sciences, Department of Physics, Guangxi University, Nanning 530004, China\\
$^2$Guangxi Key Laboratory for Relativistic Astrophysics, the Department of Physics, Guangxi University, Nanning 530004, China\\
$^3$Department of Astronomy and Institute of Theoretical Physics and Astrophysics, Xiamen University, Xiamen, Fujian 361005, China\\
$^4$College of Physics and Electronic Engineering, Nanyang Normal University, Nanyang, Henan 473061, China\\
$^5$Department of Physics and Astronomy, University of Nevada, Las Vegas, NV 89154, USA\\}
\begin{document}

\date{}

\maketitle

\begin{abstract}
The emission from black hole binaries (BHBs) and
active galactic nuclei (AGNs) displays significant aperiodic variabilities.
The most promising explanation for these variabilities is
the propagating fluctuations in the accretion flow.
It is natural to expect that the mechanism driving variabilities in BHBs and AGNs may operate
in a black hole hyper-accretion disk,
which is believed to power gamma-ray bursts (GRBs).
We study the variabilities of jet power in GRBs based on the model of propagating fluctuations.
It is found that the variabilities of jet power and the temporal profile of erratic spikes
in this scenario are similar to those in observed
light curves of prompt gamma-ray emission of GRBs.
Our results show that the mechanism driving X-ray variabilities in BHBs and AGNs may operate in the central engine
to drive the variabilities of GRBs.
\end{abstract}

\begin{keywords}
accretion, accretion disks --- gamma-ray burst: general --- galaxies: jets --- neutrinos --- X-rays: binaries
\end{keywords}

\section{Introduction}
The prompt emission of gamma-ray bursts (BHBs) are known to be highly variable.
The temporal structure exhibits diverse morphologies (\citealp{Fishman1995}),
which can vary from a single smooth large pulse to an extremely complex light curve with many erratic spikes
(see Figure 1 in \citealp{Peer2015}).
From the observational point of view,
most of the observed light curves are favored the later
case, i.e.,
presented with many spikes overlapping within a short duration (e.g., Figure 1, 4, and 5 in \citealp{Fishman1995}).
It is believed that the observed variabilities may provide an interesting clue to understand the nature of GRBs
(e.g., \citealp{Morsony2010}).
Several models have been proposed to explain the variabilities of the prompt emission,
such as internal shock model (\citealp{Rees1994}; \citealp{Paczynski1994}; \citealp{Kobayashi1997}; \citealp{Daigne1998};
\citealp{Bosnjak2009}), relativistic mini-jets (\citealp{Lyutikov2003}; \citealp{Yamazaki2004}; \citealp{Zhang2014}) or relativistic turbulence (\citealp{Narayan2009}; \citealp{Kumar2009}; \citealp{Lazar2009}; \citealp{Lin2013})
in a bulk relativistic jet.
Within the internal shock model,
the variabilities of prompt emission are attributed to the history of central engine activity.
Then, the variabilities from the central engine may be important for our understanding of GRBs nature.

Several scenarios for GRBs central engine have been discussed in the literature (see \citealp{Zhang2011} for a review).
The leading type of these scenarios is a stellar-mass black hole surrounded
by a hyper-accretion disk (e.g., \citealp{Narayan1992}; \citealp{Popham1999}; \citealp{Narayan2001}; \citealp{Gu2006}; \citealp{Liu2007}).
In this type, two kind of energy reservoirs are proposed to provide the jet power.
The first kind of energy reservoir is the rotational energy of black hole/accretion disk,
which can be extracted through the Blandford-Znajek/Blandford-Payne mechanism (BZ/BP model)
by the aid of large-scale magnetic field in the disk
(\citealp{Blandford1977}; \citealp{Blandford1982}; \citealp{Lei2013}; \citealp{Liu2015}).
The second kind of energy reservoir is the accretion energy in the disk carried by neutrinos and antineutrinos.
The annihilation of neutrino-antineutrino pairs outside the disk fuels a relativistic jet.
In these two energy reservoirs, the formation of jets is associated with the hyper-accretion disk.
Then, it is natural to expect that the variabilities in the hyper-accretion disk
would lead to the fluctuations of jet (\citealp{Lin2015}).
In black hole binaries (BHBs) and active galactic nuclei (AGNs),
X-ray emission from accretion disks is observed to display significant aperiodic variabilities on a broad range of timescales
(e.g., \citealp{Gilfanov2010}).
It reveals that fluctuations may appear in the accretion disk of BHBs and AGNs.
One basic question is that, what would
be the behaviour of jet in GRBs, if the mechanism
driving fluctuations in the disk of BHBs and AGNs
operates in a hyper-accretion disk.
This is the focus of our work.

For the variabilities of X-ray emission in BHBs and AGNs,
the most promising explanation is the model of propagating fluctuations
(\citealp{Lyubarskii1997}; \citealp{King2004}; \citealp{Mayer2006}; \citealp{Janiuk2007}; \citealp{Lin2012}).
In this model, the variabilities of disk emission arise from the fluctuations of mass accretion rate,
which are produced at different radii and propagate into the inner region of the disk.
The fluctuations of mass accretion rate in the inner region
would modulate the energy released in the vicinity of black hole,
where produces most of X-ray emission.
Observations have shown that the variabilities of X-ray emission are non-linear.
In addition, the root-mean-square ($\sigma_{\rm rms}$)
variability is proportional to the average flux over a wide range of time-scales
(\citealp{Uttley2001}; \citealp{Gleissner2004}; \citealp{Uttley2005}; \citealp{Heil2012}).
This indicates that short time-scale variations are modulated by those with longer time-scales,
and thus favours the model of propagating fluctuations.
In this work, we adopt the model of propagating fluctuations
to study the variabilities of jet power in GRBs.

\section{Procedures for producing light curves}
\subsection{Model of Propagating Fluctuations}
As first suggested by \cite{Lyubarskii1997},
if the viscosity parameter $\alpha$ fluctuates at each radius
on the local viscous timescale and in a spatially uncorrelated manner,
the characteristic power spectral density (PSD) of observed X-ray variabilities can be reproduced.
Following the work of \cite{Arevalo2006},
we discretize the accretion disk into a finite number of annulus.
The ratio between radii for consecutive annulus remains constant.
With this kind of geometric spacing and keeping a constant variability power of $\alpha$ for different annulus,
the characteristic shape of X-ray variabilities PSD can be produced.
Then, the fluctuations of $\alpha$ in an annulus is modelled as
\begin{equation}\label{EQ:alpha_fluctuation}
\alpha(r, t)  = {\alpha _0}[1 + bu(r, t)]^\xi,
\end{equation}
where $\alpha _0$, $b\;(<1)$, and $\xi\;(=1, 2)$ are constants,
$r$ is the radius of annulus relative to the central black hole,
and $u(r, t)$ sets the fluctuations of $\alpha(r, t)$.
The conditions of $\overline{u(r, t)}=0$ and $\rm{max}(\left| {u(r, t)} \right|)=1$ are adopted,
where $\overline{u(r, t)}$ and $\rm{max}(\left| {u(r, t)} \right|)$ are
the average value and maximum absolute value of $u(r, t)$, respectively.
In order to produce most of variability power at the viscous timescale $t_{\rm vis}(r)$,
$u(r, t)$ is modelled as random fluctuations with a Lorentzian-shaped-PSD,
\begin{equation}\label{EQ:Lorentzian_PSD}
{P_f} \propto \frac{{2Q{f_0}}}{{f_0^2 + 4{Q^2}{{(f - {f_0})}^2}}},
\end{equation}
where $f_0=1/t_{\rm vis}(r)$ is adopted and $Q=0.5$ determines the width of the Lorentzian-shape (\citealp{Arevalo2006}).
The relation of $t_{\rm vis}(r)=r^2/\nu$ is found for an accretion disk,
where $\nu$ is the the kinematic viscosity.
It should be noted that
the value of $b$ would determine the fluctuation of $\alpha$ or the value of $\Delta \alpha/\alpha$,
where $\Delta \alpha$ represents the amplitude of $\alpha$ fluctuation.
In the work of \cite{Lyubarskii1997},
the condition of $\Delta \alpha/\alpha <<1$ is adopted
in order to find the fluctuations of mass accretion rate by analytical method.
However, $\Delta \alpha/\alpha <<1$ does not the requirement of the propagating fluctuations model
(e.g., \citealp{Janiuk2007}).
Theoretically, a turbulent viscous stress,
which is believed to be arose from the magneto-hydrodynamic rotational instability (e.g., \citealp{Balbus1998}),
always shows high-amplitude variabilities as indicated by
numerical simulations (e.g.,  \citealp{Hirose2009}; \citealp{Simon2009}; \citealp{Davis2010}).
One can find this behavior in Figure~10 of \cite{Hirose2009},
of which the ratio of maximum $\alpha$ to minimum $\alpha$ can reach $\sim 7$.
This reveals that the value of $b$ can be as high as $\sim 0.75$.
In our work, we will discuss the situations with $b=0.2$, $0.5$, and $0.8$,
which can reproduce the observed PSD of disk emission in BHBs and AGNs (see discussions in Section~\ref{Sec:Results}).
For these situations, $b <<1$ or $\Delta \alpha/\alpha <<1$ does not satisfy.
Then, we present numerical studies on the fluctuations of mass accretion rate.

In the model of propagating fluctuations,
the fluctuations of $\alpha$ would produce a varying mass
accretion rate $\dot{M}$ in the inner region of disk.
The value of $\dot{M}$ in the disk can be estimated {\bf as}
(e.g., Equation~(3.15) in \citealp{Kato2008})
\begin{equation}\label{Eq:accretion rate}
\dot{M}(r,t)= 6\pi \sqrt{r}\frac{\partial }{{\partial r }}\left( {\nu \Sigma {r^{1/2}}} \right),
\end{equation}
where $\nu=\alpha{c_s}H$, $c_s$ is the sound velocity of gas, $H$ is the half thickness of the disk at radius $r$,
and the surface density $\Sigma$ satisfies
(e.g., Equation~(3.18) in \citealp{Kato2008})
\begin{equation}\label{EQ:Surface_Density}
\frac{{\partial \Sigma }}{{\partial t}} = \frac{1}{r}\frac{\partial }{{\partial r}}\left[ {3{r^{1/2}}\frac{\partial }{{\partial r}}\left( {{r^{1/2}}\nu \Sigma } \right)} \right].
\end{equation}
Equations~(\ref{Eq:accretion rate}) and (\ref{EQ:Surface_Density})
are the same as Equations~(4) and (6) in \cite{Lyubarskii1997}, respectively,
but with different symbol for surface density of disk.
As showed in \cite{Lyubarskii1997},
Equations~(\ref{Eq:accretion rate}) and (\ref{EQ:Surface_Density}) can present a better explanation
for the observed variabilities.
In addition, the results found in our work are the qualitative results.
Then, we use Equations~(\ref{Eq:accretion rate}) and (\ref{EQ:Surface_Density}) to study the variabilities of GRBs.
The outflow and gas adding into the disk,
which are not considered in Equation~(\ref{EQ:Surface_Density}),
may be present in the central engine of GRBs, BHBs, and AGNs.
However, the disk emission in BHBs and AGNs is also observed with highly variable
even above two ingredients are believed to be present in the disk.
That is to say, the outflow and gas adding into the disk do not
alter the highly variable behavior in BHBs, AGNs, and the central engine of GRBs (\citealp{Lin2012}).
In this work, we are interested in the variabilities of GRBs jet under the situation that
the mechanism driving X-ray variabilities in BHBs and AGNs operates in a hyper-accretion disk.
Then, the outflow and gas adding into the disk are not considered.
The value of $c_s$ and $H$ is associated with the advection factor
$f_{\rm adv}(=Q_{\rm adv}^-/Q_{\rm vis}^+)$ of the accretion flow,
i.e., $c_s\sim {\upsilon_\varphi}\sqrt{f_{\rm adv}}$ and $H/r\approx \sqrt{f_{\rm adv}}$,
where $Q_{\rm adv}^-$ ($Q_{\rm vis}^+$) is the advection cooling (viscous heating) in the accretion flow,
and $\upsilon_\varphi$ is the Kepler's rotation velocity of gas around black hole.
The $f_{\rm adv}\sim 1.0$ and $f_{\rm adv}\gtrsim 0.01$ are found in the advection-dominated accretion flow (ADAF; \citealp{Narayan1994}) and in the inner region of neutrino-cooling-dominated accretion flow (NDAF; \citealp{Popham1999}; \citealp{Kohri2005}), respectively.
Since the value of $H$ does not change significantly for these two cases,
we adopt $H/r=0.5$ in our calculations.
Then, the value of $c_s$ can be estimated with $c_s={\upsilon_\varphi}/2$.

\subsection{Jet Power}
In order to estimate the variabilities of jet power,
the dependence of jet power on $\dot{M}$ should be prescribed.
The dominant paradigms for jet production are outlined in the works of \cite{Blandford1977}
and \cite{Blandford1982}.
In both of the models (i.e., BZ/BP model),
the jet power $P_{\rm jet}$ can be modeled as (\citealp{Livio1999}; \citealp{Cao2002}; \citealp{Li2012}; \citealp{Lin2015})
\begin{equation}\label{EQ:Jet_Power}
{P_{{\rm{jet}}}}(t) \propto \int_{{r_{{\rm{in}}}}}^{{r_{{\rm{out}}}}} {\dot M(r,t)\epsilon(r)2\pi rdr}
\end{equation}
where $\epsilon(r) = (r/r_g)^{-\gamma}(1-\sqrt{r_{\rm in}/r})$ with $\gamma=3$
is taken to follow the radial loss rate of gravitational energy in the accretion flow,
$r_g(=2GM_{\rm BH}/c^2)$ is the Schwarzschild radius of black hole,
and $r_{\rm in}$ ($r_{\rm out}$) is the inner (outer) radius of disk involved to produce the jet.
The observations show that the launch region of jet in M87 is limited to
$r_{\rm out} \lesssim 10r_g$ (\citealp{Asada2012}; \citealp{Hada2013}).
Then, we adopt $r_{\rm out} = 10r_g$ and $r_{\rm in}=3r_g$ in our calculations.

In a NDAF, the jet may be powered by neutrino-antineutrino annihilation,
which depends on the luminosity of hyper-accretion disk.
The disk luminosity can be estimated with $\dot{M}$, $f_{\rm adv}$, and $\epsilon(r)$, i.e.,
\begin{equation}\label{EQ:Disk_Luminosity_init}
L(t) =\int_{{r_{{\rm{in}}}}}^{{r_{{\rm{out}}}}} {\dot M(r,t)\epsilon(r)2\pi r(1-f_{\rm adv})dr},
\end{equation}
where $r_{\rm out}$ is the outer radius of NDAF.
For disks with $\dot{M}$ satisfying $\dot{M}_{\rm ign} < \dot{M} < \dot{M}_{\rm trap}$,
the value of $f_{\rm adv}$ may be $\sim 0.01$ in the inner region of disk
and $\sim 1$ in the outer region of disk.
Here, $\dot{M}_{\rm ign}$ and $\dot{M}_{\rm trap}$
are the critical mass accretion rates for igniting and suppressing neutrino cooling, respectively (\citealp{Chen2007}).
Then, the dependence of $1-f_{\rm adv}$ on the radius may be modeled as $r^{-\mu}$ with $\mu\gtrsim 0$.
With $1-f_{\rm adv}\propto r^{-\mu}$ in Equation~(\ref{EQ:Disk_Luminosity_init}),
it is shown that the variabilities of $L(t)$ for different $\mu\;(\mu>0)$
do not show significant difference except for the variabilities in the shortest timescales
(see Figure 2 of \citealp{Arevalo2006}),
which does not affect our conclusions.
Then, we adopt $\mu=0$ in this work.
{\bf Thus}, the disk luminosity can be described as
\begin{equation}\label{EQ:Disk_Luminosity}
L(t) =\int_{{r_{{\rm{in}}}}}^{{r_{{\rm{out}}}}} {\dot M(r,t)\epsilon(r)2\pi rdr}.
\end{equation}
According to the results showed in \cite{Zalamea2011},
the jet power can be described as
\begin{equation}\label{EQ:Jet_Power_nu}
{P_{\rm{jet}, \nu\bar{\nu}}}(t) \propto L(t)^{9/4}
=\left[ {\int_{{r_{{\rm{in}}}}}^{{r_{\rm{out}}}} {\dot M(r,t)\epsilon(r)2\pi rdr}} \right]^{9/4}.
\end{equation}
If $\dot M$ is {\bf a} constant for different radius,
above equation is reduced to ${P_{\rm{jet}, \nu\bar{\nu}}}\propto {\dot M}^{9/4}$,
which can be found in Equation~(22) of \cite{Zalamea2011}.
In the model of propagating fluctuations,
$\dot M$ may be different for different $r$ and $t$.
In order to involve this behavior,
Equation~(\ref{EQ:Jet_Power_nu}) is introduced to approximately describe ${P_{\rm{jet}, \nu\bar{\nu}}}$.
The power-law index $k$ describing the relation of ${P_{\rm{jet}, \nu\bar{\nu}}}$ and $L$
as ${P_{\rm{jet}, \nu\bar{\nu}}}\propto L^k$
may be different in different work
(such as, $k=2.17$ is found in \citealp{Xue2013}).
However, a minor difference in $k$ does not play significantly effect on the variabilities (see the discussion in end of Section~\ref{Sec:Results}).
As showed by \cite{Arevalo2006}, the variabilities of ${P_{\rm{jet}, \nu\bar{\nu}}}(t)$
produced with Equation~(\ref{EQ:Jet_Power_nu}) and
constant mean value of $\dot{M}(r,t)$ do not present significantly difference
from the variabilities of $\dot{M}(r_{\rm in},t)^{9/4}$.
To simplify, Equation~(\ref{EQ:Jet_Power_nu}) is used with $r_{\rm in}=3r_g$ and $r_{\rm out}=10r_g$,
which does not affect our qualitative conclusions in this work.
Equations~(\ref{EQ:Jet_Power}) and (\ref{EQ:Jet_Power_nu})
are used to estimate the variabilities of jet power.
Since we are interested in the variabilities of jet power,
the jet power is normalized with its maximum value in our simulations.

\section{Results}\label{Sec:Results}
The black hole-torus engine is widely discussed in both the collapsar scenario
(\citealp{Woosley1993}; \citealp{MacFadyen1999}; \citealp{Proga2003}; \citealp{Zhang2003})
and the compact star merger scenario (\citealp{Narayan1992}).
In the collapsar scenario, the accretion of the stellar core,
which forms an accretion disk around the central black hole,
fuels the prompt emission.
In the compact star merger scenario,
a debris disc is formed within only a few dynamical time-scales
when the lighter companion neutron star is tidally disrupted.
Then, we discuss the variabilities of jet driven by the accretion of a torus.
The initial distribution of gas in the torus is simply described as
\begin{equation}\label{EQ:Torus_Structure}
\Sigma (r) = {\Sigma _0}\exp \left[ { - {{\left( {\frac{{r - {r_0}}}{{{r_0}/4}}} \right)}^2}} \right],
\end{equation}
where $r_0=100r_g$ is the central radius of gas distribution in the disk.
The mass of gas in the disk is set as $M_{\odot}$ (solar mass), i.e.,
$\int_{r_{\rm in}}^{\infty}{2\pi r\Sigma dr}=M_{\odot}$,
and $M_{\rm BH}=10M_{\odot}$ is adopted.

We study the variabilities of jet power in four cases:
(a) $b=0.2$, $\xi=1$; (b) $b=0.5$, $\xi=1$; (c) $b=0.8$, $\xi=1$; (d) $b=0.8$, $\xi=2$.
The varying $\alpha(10r_g,t)$ in different situations is showed in the upper panels of Figure~1.
In a magnetohydrodynamic simulation (e.g., \citealp{Hirose2009}),
the total duration of simulations may be around $500/\Omega_{\rm K}$,
where $\Omega_{\rm K}$ is the orbital frequency.
For performing comparison, we show the variabilities of $\alpha(10r_g,t)$ in $[0{\rm s}, 2.5{\rm s}]$,
where $2.5{\rm s }$ is around $500/\Omega_{\rm K}(10r_g)$.
It can be found that
the variabilities of $\alpha$ (such as the ratio of maximum $\alpha$ to minimum $\alpha$) showed in the upper panels,
especially for cases (b) and (c),
are similar to those found in the simulations
(e.g., Figure~10 of \citealp{Hirose2009}).
In the middle and lower panels of Figure~1,
we show the disk emission ($L_{\rm X}$) of BHBs and the corresponding PSD, respectively.
\begin{figure*}
\centering
\includegraphics[width=\linewidth]{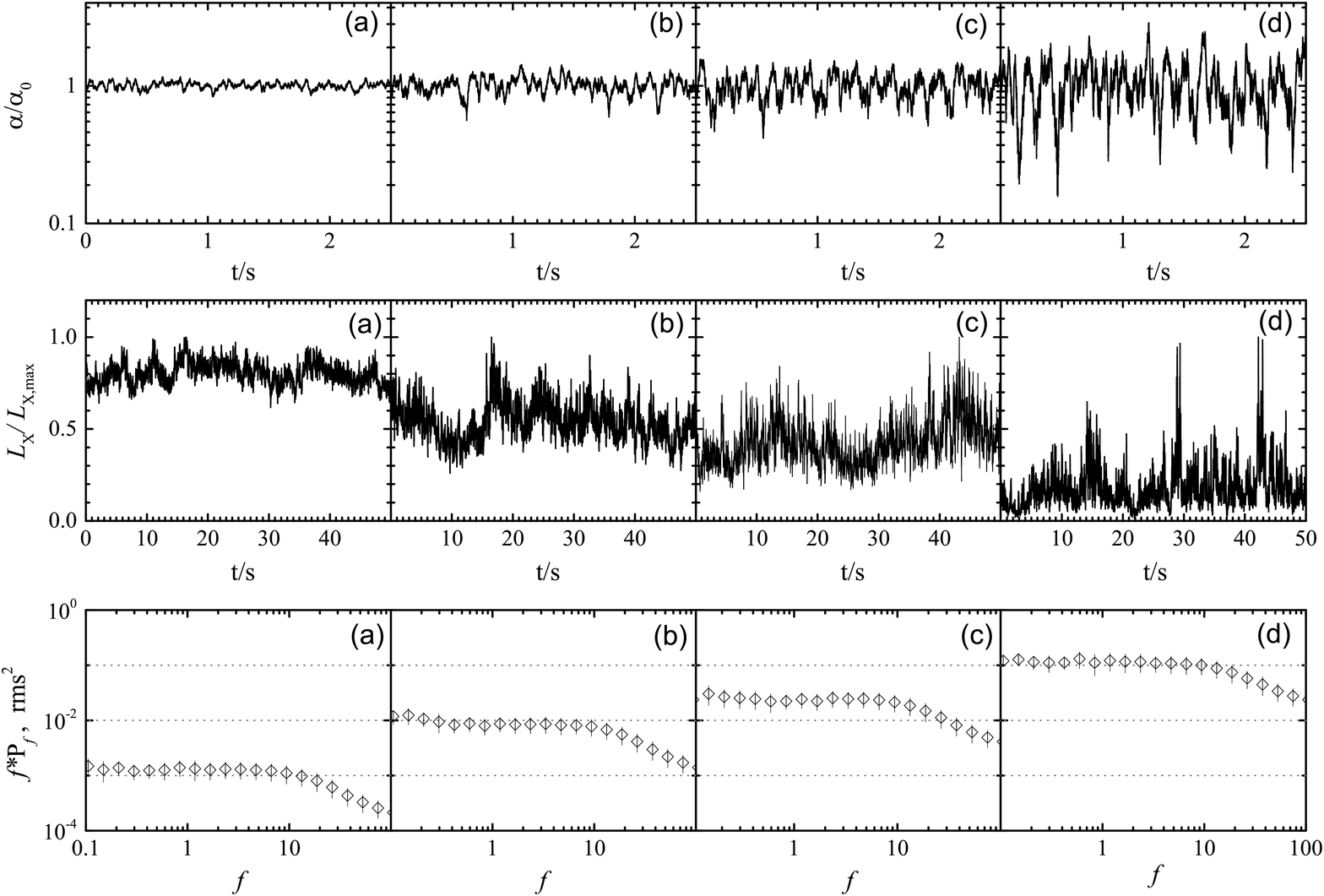}
\caption{Variabilities of $\alpha(10r_g,t)$ (upper panels), the light curves of $L_{\rm X}$ from BHBs (middle panels),
and the PSD of $L_{\rm X}$ (lower panels) for cases (a), (b), (c), and (d), respectively.}
\label{Fig:Torus_Accretion_10Rg}
\end{figure*}
Here, $L_{\rm X}\propto \int_{r_{\rm in}}^{100r_g} {\dot M(r,t)^2\epsilon(r)2\pi rdr}$
with $\gamma=5$ in $\epsilon$ is adopted to describe
the X-ray emission from ADAF ($\dot M\lesssim\dot 0.01M_{\rm Edd}$),
and is calculated in a disk with a constant mass accretion rate at the outer radius ($10^3r_g$) of disk.
In the middle panels, the light curves of $L_{\rm X}$ are showed with a duration of 50s and
$L_{\rm X,max}$ is the maximum value of $L_{\rm X}$,
where the total duration of disk emission from our simulations is around $10^3\rm s$.
The light curves of $L_{\rm X}$ in the situations of (a)-(c) are always observed in BHBs.
The case (d) is used to discuss the situation with intense variabilities,
which may appear in the hard-to-soft state transition of BHBs (\citealp{Remillard2006}).
In the lower panels of Figure~1, the behavior of $f\times P_f=\rm constant$ can be found for low frequency $f<10\rm Hz$.
This reveals $P_f\propto f^{-1}$ for low frequency,
which is consistent with the expectation of propagating fluctuations model (see Figure 2 of \citealp{Arevalo2006}).
By comparing our PSD with those showed in observations (e.g., Figure 2.6 of \citealp{Gilfanov2010}),
the value of $b\sim 0.2-0.8$ is required in order to explain the observations of BHBs and AGNs.
Then, our studying four cases may all appear in the observations of BHBs and AGNs.

The light curves of jet power in the above four cases are showed in Figure~2.
In this figure,
the upper panels describe the light curves of jet power produced by BZ/BP model, i.e., Equation~(\ref{EQ:Jet_Power}),
and the lower panels describe the light curves of jet power produced
by neutrino-antineutrino annihilation, i.e., Equation~(\ref{EQ:Jet_Power_nu}).
It can be found that the light curves in Figure~2 are similar to those of observations (e.g., Figure 1.3 of \citealp{Bouvier2010}).
Moreover, the profile of erratic spikes in (b)-(d) is also similar to those of observations,
such as the erratic spikes in GRBs 920513 and 050117.
The temporal structure of jet power variabilities is
related to the value of $b$ or the amplitude of $\alpha$ fluctuations according to this figure.
If the amplitude of $\alpha$ fluctuations is high, e.g., case (d),
the light curve of jet power is found to be complex with large amplitude erratic spikes.
However, if $\alpha$ is present with low amplitude variabilities, e.g., case (a),
the light curve of jet power is smooth.
Then, one can conclude that by changing the amplitude of $\alpha$ fluctuations,
the diverse morphologies of observed light curves from a single smooth large pulse
to extremely complex light curves can be reproduced in the model of propagating fluctuations.
Making comparison between upper and lower panels in Figure~2,
one can find that the amplitude of $P_{\rm{jet}, \nu\bar{\nu}}(t)$ variabilities is
higher than that of $P_{\rm jet}(t)$ variabilities.
This is owing to the relations of
$P_{\rm{jet}, \nu\bar{\nu}}(t)\propto \dot{M}^{2.25}$ and $P_{\rm jet}(t)\propto \dot{M}$,
or, $\delta{P_{\rm{jet}, \nu\bar{\nu}}(t)}/{P_{\rm{jet}, \nu\bar{\nu}}(t)}\sim 2.25\delta{\dot{M}}/\dot{M}$
and $\delta{P_{\rm jet}(t)}/{P_{\rm jet}(t)}\sim \delta{\dot{M}}/\dot{M}$,
where $\delta{P_{\rm{jet}, \nu\bar{\nu}}}$, $\delta{P_{\rm jet}}$, and $\delta{\dot{M}}$ represent the variable component
of $P_{\rm{jet}, \nu\bar{\nu}}$, $P_{\rm jet}$, and $\dot{M}$, respectively.
According to the above two relations, the same $\delta{\dot{M}}/\dot{M}$ would produce different amplitude of $P_{\rm{jet}, \nu\bar{\nu}}(t)$ and $P_{\rm jet}(t)$ variabilities.
\begin{figure*}
\centering
\includegraphics[width=\linewidth]{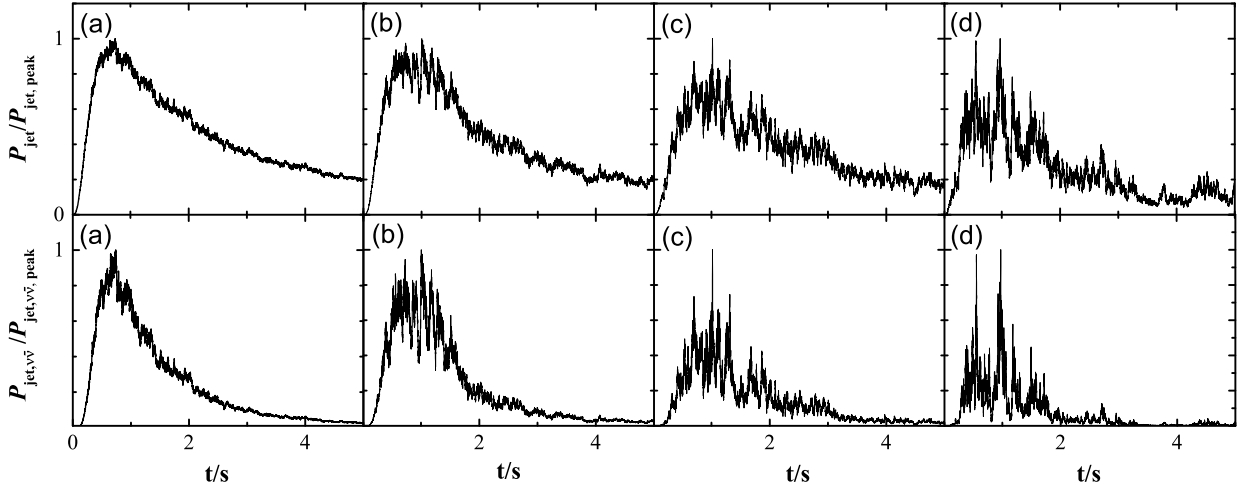}
\caption{Light curves of jet power for the cases (a), (b), (c), and (d), respectively.
The upper and lower panels describe the light curves of jet power produced by BZ/BP model and
by neutrino-antineutrino annihilation, respectively.}
\label{Fig:Torus_Accretion_10Rg}
\end{figure*}

\section{Conclusions and Discussion}
It is found that the X-ray emission from accretion disk in BHBs and AGNs displays significant aperiodic variabilities.
The most promising explanation for these variabilities is the propagating fluctuations in the accretion flow.
In addition, it is believed that a hyper-accretion disk may reside in the central engine of GRBs.
Then, it is natural to expect that the mechanism (e.g., the model of propagating fluctuations)
driving variabilities of X-ray emission in BHBs and AGNs may operate in a hyper-accretion disk.
In this paper, we study the variabilities of jet power in GRBs based on the model of propagating fluctuations.
It is found that the variabilities of jet power and the profile of erratic spikes in this scenario are similar to those
found in the prompt gamma-ray emission of GRBs.
The temporal structure of jet power variabilities is
related to the amplitude of $\alpha$ fluctuations.
If the amplitude of $\alpha$ fluctuations is high,
the light curve of simulated jet power is complex with large amplitude erratic spikes.
However, if the amplitude of $\alpha$ fluctuations is low,
the light curve of jet power is smooth.
Then, we conclude that by changing the amplitude of $\alpha$ fluctuations,
the diverse morphologies of light curves from a single smooth large pulse
to an extremely complex light curve can be reproduced in the model of propagating fluctuations.
These qualitative results reveal that
the mechanism driving X-ray variabilities in BHBs and AGNs may operate in the central engine of GRBs
to drive the observed variabilities (\citealp{Carballido2011}).

In GRB physics, the central engine of GRBs remains an open question (\citealp{Zhang2011}).
Except for scenario of a stellar-mass black hole surrounded by a hyper-accretion disk,
other scenarios have been proposed to drive GRBs.
Millisecond magnetar,
which is a rapidly spinning and strongly magnetized neutron star,
has been suggested as the central engine of GRBs
(\citealp{Usov1992}; \citealp{Thompson1994}; \citealp{Dai1998}; \citealp{Wheeler2000};
\citealp{Zhang2001}; \citealp{Metzger2008, Metzger2011}; \citealp{Bucciantini2012}; \citealp{Lv2014}).
In this scenario, the rotational energy of the millisecond magnetar is used to power GRBs
and the hyper-accretion disk is absent.
Then, the conclusion in this work,
which is based on a black hole hyper-accretion disk,
is not applicable for this scenario.

\section*{Acknowledgments}
We thank Bing Zhang, Zi-Gao Dai, and Shi-Fu Zhu for helpful suggestions.
This work is supported by the National Basic Research Program of China (973 Program, grant No. 2014CB845800),
the National Natural Science Foundation of China (Grant Nos. 11403005, 11533003, 11503011, 11573023, 11333004, 11473022, U1331101),
the Guangxi Science Foundation (Grant Nos. 2014GXNSFBA118004, 2014GXNSFBA118009, 2013GXNSFFA019001),
the Project Sponsored by the Scientific Research Foundation of Guangxi University (Grant No. XJZ140331),
and the Scientific Research Foundation of Nanyang Normal University (Grant No. ZX2015001).

\bsp

\label{lastpage}

\end{document}